\documentclass[aps,prl,twocolumn,superscriptaddress,showpacs,amsfonts]{revtex4}
\usepackage{epsfig}
\usepackage{graphicx,graphics}
\usepackage{amsmath,ulem}
\usepackage{amssymb}
\usepackage{color}
\usepackage{bm}

\begin{document}

\title{Magnetic intragap states and mixed parity pairing at
the edge of spin triplet superconductors}
\author{Alfonso Romano}
\affiliation{CNR-SPIN, I-84084 Fisciano (Salerno), Italy and
Dipartimento di Fisica ``E. R. Caianiello'', Universit\`a di
Salerno, I-84084 Fisciano (Salerno), Italy}

\author{Paola Gentile}
\affiliation{CNR-SPIN, I-84084 Fisciano (Salerno), Italy and
Dipartimento di Fisica ``E. R. Caianiello'', Universit\`a di
Salerno, I-84084 Fisciano (Salerno), Italy}
\author{Canio Noce}
\affiliation{CNR-SPIN, I-84084 Fisciano (Salerno), Italy and
Dipartimento di Fisica ``E. R. Caianiello'', Universit\`a di
Salerno, I-84084 Fisciano (Salerno), Italy}
\author{Ilya Vekhter}
\affiliation{Department of Physics and Astronomy, Louisiana State
University, Baton Rouge, Louisiana, 70803, USA}
\author{Mario Cuoco}
\affiliation{CNR-SPIN, I-84084 Fisciano (Salerno), Italy and
Dipartimento di Fisica ``E. R. Caianiello'', Universit\`a di
Salerno, I-84084 Fisciano (Salerno), Italy}
%\date{\today}

\begin{abstract}
We show that a spontaneous magnetic moment may appear at the edge
of a spin-triplet superconductor if the system allows for pairing
in a subdominant channel. To unveil the microscopic mechanism
behind such effect we combine numerical solution of the
Bogoliubov-De Gennes equations for a tight-binding model with
nearest-neighbor attraction, and the symmetry based
Ginzburg-Landau approach. We find that a potential barrier
modulating the electronic density near the edge of the system
leads to a non-unitary superconducting state close to the boundary
where spin-singlet pairing coexists with the dominant triplet
superconducting order. We demonstrate that the spin polarization
at the edge appears due to the inhomogeneity of the non-unitary
state and originates in the lifting of the spin-degeneracy of the
Andreev bound-states.
\end{abstract}


\maketitle

%\section{Introduction}

{\it Introduction.} Recognition that the surface states in
correlated materials reflect the nature of the interactions and
orders in the bulk has led to a significant research effort aimed
at the understanding, and potential control, of these electronic
states~\cite{Eschrig:PT,TopoReviews,Tanaka11}. Gapless modes at
the boundary of materials whose bulk is gapped are especially
interesting since the surface states are robust, and may be
topologically protected, i.e. their existence relies on the global
symmetries of the bulk state and does not depend on the details of
the surface scattering and other sample-dependent
parameters~\cite{TopoReviews}. The bulk gap may be due to the band
structure, or, in a metal, may arise from electron-electron
interaction, as in superconductors~\cite{d-wave}. Simple band
insulators or conventional superconductors do not support robust
low-energy states at the boundary. It is the study of their
counterparts, where the bulk is topologically non-trivial, and
hence the bulk-boundary correspondence theorem dictates the
existence of the surface states, that has been a focus of much
recent attention~\cite{Eschrig:PT,TopoReviews,d-wave}.

A prime candidate for the topological superconductivity is
Sr$_2$RuO$_4$, where the emergent consensus indicates triplet
chiral pairing, with time-reversal symmetry broken by the orbital
degrees of freedom~\cite{Maeno_RMP}. In this material signatures
of the predicted topologically protected edge states were recently
found in tunnelling spectroscopy~\cite{Kashiwaya11}. The
quasiparticles reflecting off the sample boundary experience the
sign change of the superconducting order parameter along their
trajectory, which gives rise to so-called Andreev bound states
(ABS) near the surface, which contribute to Josephson
currents~\cite{Barash}. Emergence of ABS has been investigated in
high-T$_c$ cuprates and other unconventional
superconductors~\cite{d-wave,Sauls1,LGreene}.

In this Letter we investigate the nature of the Andreev bound
states at the surface of spin triplet superconductors. We perform
a microscopic self-consistent calculation, and include a realistic
surface barrier of finite width and height, and the possibility of
pairing in one or more subdominant channels~\cite{Sauls1}. We find
that a) a subdominant {\sl in-phase} $s$-wave superconducting
order exists near the edge of the sample; b) the in-phase $s$-wave
component gives a non-unitary superconducting state at the
boundary; c) as a result, the ABS are spin-polarized, leading to a
finite surface magnetization; d) spin current flows along the
interface in this regime; e) surface charge currents exhibit
anomalous dependence on the magnetization. We analyze the
conditions for the existence of the magnetic surface states, and
investigate their spectrum numerically. These results are
supported by symmetry analysis of the Ginzburg-Landau expansion of
the free energy. Our work strongly suggests that triplet
superconductors can be used in spin-active heterostructures.

{\it Model and formalism.} We consider a two-dimensional
superconductor in a parallel slab geometry  in vacuum. If $x$ and
$y$ are the directions perpendicular and parallel to the
interfaces, respectively, the system is uniform along the $y$
axis, so that the translational symmetry is broken only in the $x$
direction. The Hamiltonian is then defined on a square lattice of
size $L\times L$ (the lattice constant is unity), with periodic
boundary conditions along $y$,
\begin{eqnarray} H&=&
-t\sum_{\langle \mathbf{i} ,\mathbf{j} \rangle,\,\sigma}
(c^{\dagger}_{\mathbf{i}\,\sigma}
c_{\mathbf{j}\,\sigma}+\text{h.c.}) -\mu \sum_{\mathbf{i},\sigma}
n_{\mathbf{i}\sigma} \nonumber \\
&&
- \sum_{\langle \mathbf{i} ,\mathbf{j} \rangle} V\left( n_{\mathbf{i} \uparrow}
n_{\mathbf{j}\downarrow}+n_{\mathbf{i}\downarrow} n_{\mathbf{j}
\uparrow} \right) + \, \sum_{\mathbf{i}} U(i_x) n_{\mathbf{i}\sigma} \; .
\label{eq:H}
\end{eqnarray}
Here the lattice sites are labelled by
$\mathbf{i}\equiv(i_x,i_y)$, with $i_x$ and $i_y$ integers between
$0$ and $L$, $\langle \mathbf{i},\mathbf{j}\rangle $ denote
nearest-neighbor sites, and $\mu$ is the chemical potential. The
nearest-neighbor attractive interaction $-V$ $(V>0)$ is effective
in both singlet and triplet pairing channels. All the energies are
in units of the hopping parameter $t$. The slab edges are located
at $i_x=0$ and $i_x=L$, and we introduce a site-dependent
potential $U(i_x)$ to model the interface barrier. To investigate
the model of Eq.~\eqref{eq:H} we decouple the interaction term in
the Hartree-Fock approximation by introducing the pairing
amplitude on a bond, $\Delta_{\mathbf{i}\mathbf{j}}=\langle
c_{\mathbf{i}\,\uparrow} c_{\mathbf{j}\,\downarrow} \rangle$, so
that  $V n_{\mathbf{i}\uparrow} n_{\mathbf{j}\downarrow}\simeq
V(\Delta_{\mathbf{i}\mathbf{j}}
c^{\dagger}_{\mathbf{j}\,\downarrow}
c^{\dagger}_{\mathbf{i}\,\uparrow}+\Delta^{*}_{\mathbf{i}\mathbf{j}}
c_{\mathbf{i}\,\uparrow}
c_{\mathbf{j}\,\downarrow}-|\Delta_{\mathbf{i}\mathbf{j}}|^2)$.
These pairing amplitudes yield the spin singlet ($S$) and triplet
($T$) components,
$\Delta^{S,T}=(\Delta_{\mathbf{i}\mathbf{j}}\pm\Delta_{\mathbf{j}\mathbf{i}})/2$,
that define the superconducting order parameters (OPs) with $s$-
or $p$-wave symmetry, i.e.
$\Delta_{s}(\mathbf{i})=(\Delta^{S}_{\mathbf{i},\mathbf{i}+\text{\^{x}}}+\Delta^{S}_{\mathbf{i},\mathbf{i}-\text{\^{x}}}+
\Delta^{S}_{\mathbf{i},\mathbf{i}+\text{\^{y}}}+\Delta^{S}_{\mathbf{i},\mathbf{i}-\text{\^{y}}})/4$
and
$\Delta_{p_{x(y)}}(\mathbf{i})=(\Delta^{T}_{\mathbf{i},\mathbf{i}+\text{\^{x}}(\text{\^{y}})}-\Delta^{T}_{\mathbf{i},\mathbf{i}-\text{\^{x}}(\text{\^{y}})})/2$,
which are then determined self-consistently~\cite{Cuoco08}.
Singlet $d$-wave superconductivity is possible but does not appear
in the parameter range where we work~\cite{Cuoco08}. In the bulk
($U=0$) the most favorable pairing state for this model depends on
the electron density, $n$, and  the chiral $p_x+ip_y$ order is
stabilized in the region between half-filling, $\mu\simeq0$, and
high (low) density ($|\mu|\simeq 2.5)$~\cite{Kuboki01}. Hence, we
choose $|\mu|\simeq 1.8$, in this window of stability, so that for
$U=0$ the filling is $n\approx0.4$. All the numerical results
below have been obtained for a pairing interaction $V=2.5$, a
rectangular potential barrier of height $U$ near the left edge of
the system, $0\leq i_x\leq 8$, and a system size $L=80$; greater
values of $L$ leave the results qualitatively unchanged. We also
determine the local spin and charge currents,
$J_{s}(i_x)=J_{\uparrow}(i_x)-J_{\downarrow}(i_x)$, and, $J_c
(i_x)=J_{\uparrow}(i_x)+J_{\downarrow}(i_x)$, with
$J_{\sigma}(i_x)=\frac{2t}{L_y} \sum_{k_y} sin(k_y) \langle
c^{\dagger}_{k_y \sigma} c_{k_y \sigma} \rangle $~\cite{Kuboki04}.
We find qualitative differences between our results for the
extended barrier, and those obtained assuming a sharp step-like
potential at the surface~\cite{BTK}. One crucial distinction is
that a finite-width barrier changes the electron density near the
boundary, and, if it is strong enough to drive the density into
the regime where superconducting components competing with the
dominant triplet order are stabilized, leads to the coexistence of
two distinct pairing states near the interface. This coexistence
is at the root of the phenomena we describe.
%enabling the emergence of
%once the potential is strong enough to

%%%%%%%%%%%%%%%%%%%%%%%%%%%%%%%%%%%%%%%%%%%%%%%%%%%%%%%%%%%%%%%
%
\begin{figure}[t]
\begin{center}
\includegraphics[width=0.43\textwidth]{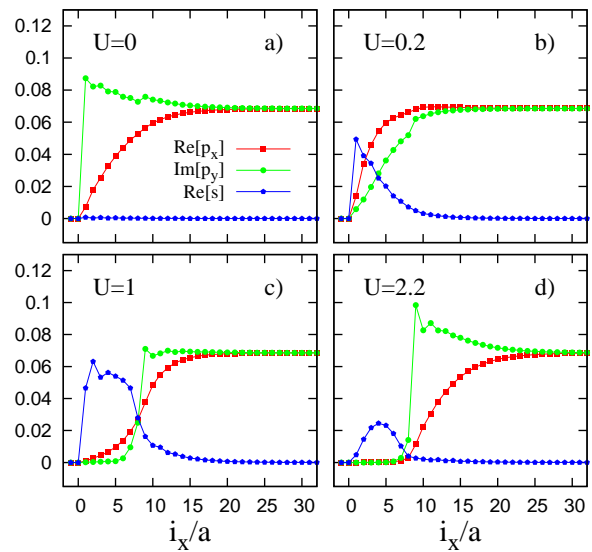}
\end{center}
\caption{(color online) Evolution of the order parameters for
different heights of the potential barrier $U$ extending from
$i_x=0$ to $i_x=8$. Label $s$-wave refers to nearest-neighbor
pairing; Note that the $s$-wave amplitude is purely
real.}\label{OP_mag}
\end{figure}
%
%%%%%%%%%%%%%%%%%%%%%%%%%%%%%%%%%%%%%%%%%%%%%%%%%%%%%%%%%%%%%%%

%%%%%%%%%%%%%%%%%%%%%%%%%%%%%%%%%%%%%%%%%%%%%%%%%%%%%%%%%%%%%%%
%
\begin{figure}[t]
\begin{center}
\includegraphics[width=0.43\textwidth]{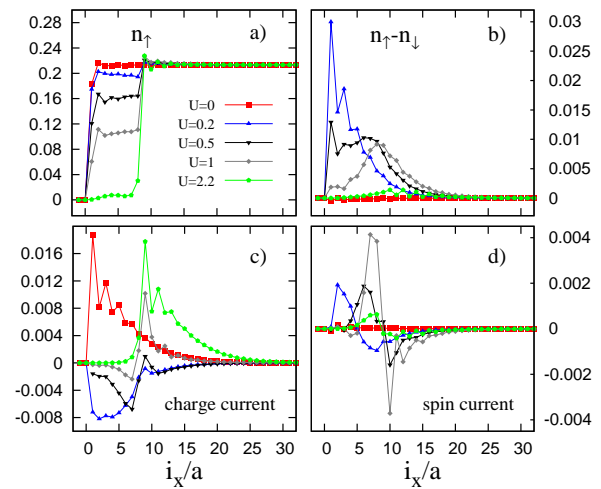}
\end{center}
\caption{(color online) Evolution of the majority spin electron
density: (a), spin-polarization (b), as well as of the spin (c)
and charge currents (d) for several values of the potential $U$.
Note the sign change of the charge current in the regime where the
magnetization appears. Also, the spin polarization and the spin
current are peaked in the range where the superconducting orders
of different symmetry coexist, cf. Fig.~1.}\label{current}
\end{figure}
%
%%%%%%%%%%%%%%%%%%%%%%%%%%%%%%%%%%%%%%%%%%%%%%%%%%%%%%%%%%%%%%%

{\it Numerical results.} Figs.\ref{OP_mag},\ref{current} show
representative results for the electron density, spin
polarization, spin and charge currents, as well as the evolution
of the superconducting order parameters for different strength of
the surface potential. For $U=0$, Fig.\ref{OP_mag}(a), we find the
expected result: the interface is pairbreaking for the $p_x$
component of the OP, while the $p_y$ component remains essentially
constant. Finite $U$ depletes the electron density near the edge,
Fig.\ref{current}(a), and, as $U$ exceeds a critical magnitude,
here found to be $U_c\simeq 0.19$, the surface electron density
reaches the value where a subdominant $s$-wave component of the
order parameter first appears, Fig.\ref{OP_mag}(b)-(d).
Consequently, there is a substantial region of coexistence of the
superconducting OPs with different parity. Note that mixed parity
is allowed here since the presence of the barrier breaks the
inversion symmetry. Remarkably, the emergence of the mixed-parity
phase is accompanied by the appearance of a finite spin
polarization in that same region, Fig.\ref{current}(b), as well as
that of a spin current, Fig.\ref{current}(d). At the same time the
surface charge current, initially present simply due to the chiral
nature of the bulk superconducting state~\cite{Maeno_RMP}, changes
sign, Fig.\ref{current}(c). As the barrier height increases, fewer
carriers remain in the boundary layer, and the magnetization and
other signatures of the unconventional surface states gradually
disappear. Note that the charge currents at the interface give
rise to a magnetic field, and, therefore, to a spin
polarization~\cite{MineevBook,Imai2012}. However, as discussed in
the supplementary material ~\cite{suppl}, the origin and the
spatial profile of this field are very different from those of the
spin polarization shown in Fig.\ref{current}(b).
%charge-currents a magnetic field is also induced. For the present
%two-dimensional slab, according to the Maxwell's law $\nabla\times B=\mu_0 J_c$, the magnetic field is perpendicular to the
%plane ($z$-direction). At a distance $i_x$ from the edge it has an
%amplitude $B_{orb}^z(i_x)=\mu_0 \sum_{j_x=1}^{i_x}
%J_c(j_x)$~\cite{Imai2012}. For $U=0$ there is an induced magnetic
%field but no net spin magnetization, while above the threshold
%both contributions occur (Fig.\ref{current}(c)). As expected, due
%to the different generating mechanisms, the spatial profile of the
%induced magnetic field is not concord ant with that of the
%magnetization. It changes sign as a consequence of the sign
%variation of the charge current. In doing that we have neglected
%the Meissner screening that would have introduced counter currents
%suppressing this field on length scale that are greater than the
%London penetration depth (inner side of the superconductor). When
%converting the magnetization into a magnetic field intensity, its
%order of magnitude is comparable with that of $B_{orb}$ due to the
%edge charge currents~\cite{Imai2012}.}

\begin{figure}[!t]
\begin{center}
\includegraphics[width=0.43\textwidth]{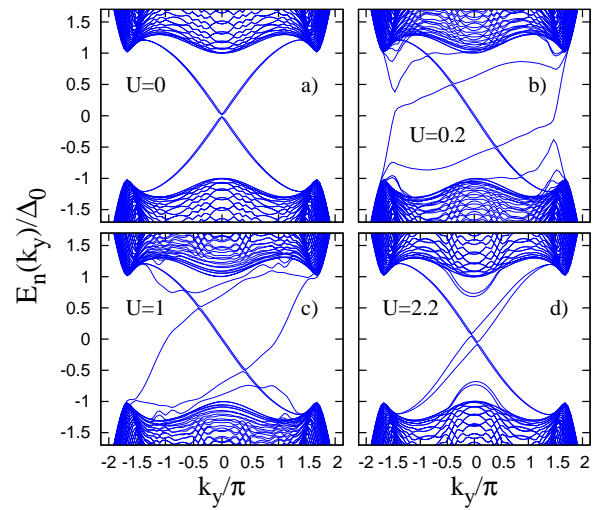}
\end{center}
\caption{(color online) Energy spectrum of the solutions of the
Bogoliubov-de Gennes equations scaled to the superconducting gap
$\Delta_0$ of the homogeneous spin triplet superconductor. a) pure
chiral phase. Two dispersing branches crossing $E=0$ correspond to
the modes along the left and the right edges of the sample. b)-d):
mixed-parity phases. The right edge mode remains unchanged while
the modes along the left edge become spin-split. Extra midgap
states which are spin-polarized appear at the bottom (occupied)
and the top edge (unoccupied) of the gap reflecting the occurrence
of a mixed-parity coexisting region. The latter and the spin-split
chiral states both contribute to the overall spin accumulation.
}\label{spectrum}
\end{figure}

The analysis of the energy spectrum $E_n(k_y)$ (due to
translational invariance along the interface $k_y$ is a good
quantum number) obtained from the numerical solution of the
Bogoliubov-De Gennes equations confirms that the local
magnetization is due to the gapless modes propagating in one
direction along the boundary. This is evident from
Fig.~\ref{spectrum}, where the two originally spin-degenerate
chiral edge states (Fig.~\ref{spectrum}(a)) associated with the
left boundary split once the barrier potential exceeds $U_c\simeq
0.19$ (Fig.~\ref{spectrum}(b)-(d)). This splitting appears
together with the spin polarization at that edge (of course, the
counterpropagating mode at the right edge is not affected). It is
clear, for example, from Fig.~\ref{spectrum}(b), that splitting
leads to the unequal number of occupied ($E_n(k_y)<0$) states for
the two spin-split modes. In addition, new intragap states appear
close the bottom and the top of the gap edge. They evolve from an
asymmetric band (peak at $k_y>0$) for $U=0.2$ and $1.0$
(Fig.~\ref{spectrum}(b)-(c)), when singlet-triplet coexistence is
significant, to a symmetric ``bump'' around $k_y=0$ for
$U=2.2$(Fig.~\ref{spectrum}(d)). This feature contributes
significantly to the spin current and is in large part responsible
for the change in sign of the charge current. Note that the
branches crossing the gap are piecewise-linear, so that
$J_{\sigma}\propto\sum_{\bm k} dE_\sigma/dk \Theta(-E_{\sigma,\bm
k})$ nearly compensates between the spin-split central branches.
As the barrier height $U$ increases the splitting between the edge
modes at the edge diminishes, and magnetization is reduced as the
boundary region becomes depleted.

% with a structure that evolves from an asymmetric
%profile in the case of an extended region of singlet-triplet
%coexistence at the surface, i.e. for $U=0.2$ and $1.0$, to a more
%symmetric one around $k_y=0$ for $U=2.2$. This asymmetric ``bump''
%contributes significantly to the spin current and is in large part
%responsible for the change in sign of the charge current. As the barrier height is increased
%the splitting between the spin-up and spin-down modes at the edge
%decreases, and eventually leads to lower and lower values of the
%magnetization, as the boundary region becomes depleted.

Our result shows that a spin accumulation may occur in a triplet
superconductor {\it without the proximity coupling to an exotic
system}. This situation is quite different from the case discussed
in Refs.\onlinecite{Sengupta08,Lu09,Yang11}, where spin
polarization appears at the sharp interface between semi-infinite
triplet and a singlet superconductors with no coexistence region,
when analyzed using the BTK scattering formulation~\cite{BTK}. In
Refs.~\onlinecite{Sengupta08,Lu09,Yang11} such magnetization
appears only for a non-trivial phase difference between the two
superconductors, exactly the opposite to what we find in our
geometry.

The essential ingredient of our finding is the coexistence of the
singlet and triplet OPs in the same material over a finite range
near the interface, obtained from a fully self-consistent solution
of the Bogoliubov-de Gennes equations in the presence of a
realistic boundary potential. In our case the magnetization exists
near the surface over the same length scale as the density
gradient: this is because for our parameter values the
superconducting coherence length, $\xi_0$, is comparable to that
width. If screening of the surface potential occurs on a shorter
scale, the surface barrier nucleates the subdominant $s$-wave
component, and the coexistence range is set by the coherence
length.
%%%%%%%%%%%%%%%%%%%%%%%%%%%%%%%%%%%%%%%%%%%%%%%%%%%%%%%%%%%%%%%
%
%
%%%%%%%%%%%%%%%%%%%%%%%%%%%%%%%%%%%%%%%%%%%%%%%%%%%%%%%%%%%%%%%

{\it Symmetry analysis.} To elucidate the origin of the spin
polarization we analyse the problem from the symmetry perspective.
The main insight from the numerical results is that the singlet
and the triplet OPs coexist near the edge due to the finite range
of the boundary potential, and the absence of the inversion
symmetry. We first note that in this situation the OP is
non-unitary. In the 2x2 spin space the order parameter is
~\cite{SigUeda} ${\bm\Delta}_{\bm k}=i\left[\left(\bm d_{\bm
k}\cdot\bm\sigma\right)+\psi_{\bm k}\right]\sigma_y$, where
$\sigma_i$ are Pauli matrices, and $\bm d_{\bm k}$ and $\psi_{\bm
k}$ are the three triplet and the singlet pairing amplitudes
respectively. The gauge-invariant product is given by
${\bm\Delta}{\bm\Delta}^\dagger = |\bm d_{\bm k}|^2\sigma_0+
|\psi_{\bm k}|^2\sigma_0+\bm q_{\bm k}\cdot\bm\sigma $ with $\bm
q_{\bm k}=\psi_{\bm k}\bm d_{\bm k}^\star+\psi_{\bm k}^\star\bm
d_{\bm k}+ i\left[\bm d_{\bm k}\times\bm d_{\bm k}^\star\right]$.
In our case the bulk triplet superconductor is unitary, $\bm
d\|\widehat{\bm z}$ and therefore $\bm d_{\bm k}\times\bm d_{\bm
k}^\star=0$. However, in the coexistence region, $\bm q_{\bm
k}=2{\rm Re}\left[\psi_{\bm k}\bm d_{\bm k}^\star\right]\neq 0$
since, as Fig.~\ref{OP_mag} shows, the self-consistent solution
yields the {\it in-phase} singlet and triplet $p_x$ components.

The net spin polarization of the Cooper pairs,
%$s_i=(1/2)Tr({\bm\Delta}{\sigma_i}{\bm \Delta}^\dagger)$ is
%generally given by
$s_i=(1/2)\sum_{\bm k} (\bm q_{\bm k}\cdot\bm\sigma) \sigma_i$,
does not appear for {\it uniform} mixed parity system (with
unitary triplet component) since the singlet and triplet order
parameters have opposite parity. Hence their product is odd in
momentum and its average over $\bm k$ vanishes. The same argument
does not hold in a system without translational invariance such as
the one considered here, and the $\bm q$-vector that is
non-uniform in space yields spin-splitting of the edge states. To
analyse this situation we consider a Ginzburg-Landau (GL) magnetic
free energy density $f_m$ in the region of coexistence. We choose
the direction of the magnetization $\bm m$ parallel to $\bm d(\bm
r)=\widehat{\bm z}\left(\eta_x(\bm r) +i \eta_y(\bm r) \right)$,
along the $\widehat{\bm z}$-axis, as required by the spin rotation
invariance, and assume that the pairing amplitudes depend only on
the coordinate $x$ normal to the boundary. The GL expansion
includes terms linear in $m$ and the gradient of the
$p_x$-component of the triplet pairing,
\begin{eqnarray}
  f_m&=& m^2+ (\partial_x m)^2+ \alpha
  m \Bigl\{\beta
  \partial_x\left[\eta_x\psi^\star+\eta_x^\star\psi\right]+
  \\
  \nonumber
  &&+
  \left[\psi^\star(\partial_x \eta_x)+\psi(\partial_x \eta_x^\star)\right]-
         \left[\eta_x^\star(\partial_x \psi)+\eta_x(\partial_x\psi^\star)\right]
\Bigr\}\,.
\label{eq:gl}
\end{eqnarray}
Such linear coupling means that in the region of the coexistence a
finite magnetization always appears unless the singlet and the
triplet $p_x$ components are out of phase (and the product $\eta_x
\psi$ is purely imaginary). This emphatically brings forth the
distinction between our results and those for the S-I-S
junction~\cite{Sengupta08,Lu09,Yang11}, where the spin
accumulation only occurs if the two pairing components are out of
phase, the exact opposite of the result we find. We note that this
term (and its analog derived in Ref.~\cite{Kuboki:11}) has a
structure which is different from the well-known contributions
that relate the inhomogeneity in $\eta(\bm{r}) \times
\eta^*(\bm{r})$ \cite{MineevBook} to spontaneous charge-currents
at the edge and a magnetic field~\cite{suppl}.

Only the $p_x$-component of the triplet appears in the GL
expansion above; in principle the term $\partial_y \eta_y$ is also
allowed by symmetry~\cite{Kuboki:11}, but vanishes under the
assumption of the translational invariance along the interface. It
follows that the time-reversal symmetry breaking by the bulk
chiral triplet state is not at the origin of the magnetization of
the Andreev bound states: the same result would be achieved for
purely real $p_x$ bulk triplet superconductivity, while for the
imaginary $p_x$ bulk pairing with real subdominant $s$-wave
pairing near the interface no magnetization appears. The results
in the supplementary material~\cite{suppl} show precisely this
behavior. Note that the terms written in Eq.~(2) do not contribute
to the currents along the interface as the only gradient is along
the $x$ direction. To obtain such currents, higher order terms
involving the $\eta_y$ component of the order parameter are
needed. For non-chiral single-component $p_x$ triplet order
parameter no substantial spin and charge currents exist, even
though the magnetization still appears. Such currents are allowed
by symmetry, but their magnitude is strongly suppressed, likely by
a power of $T_c/t$.

{\it Discussion.} We showed that Andreev bound states near a
boundary of a triplet superconductor can be spin-polarized and
yield nontrivial spin and charge currents near the interface. The
origin of the spin polarization is in the emergence of the
coexistence regime of the triplet and the subdominant singlet
pairing components near the boundary above a critical surface
barrier. Our numerical results demonstrate that the two are
phase-locked, and both the numerical fully self-consistent
solution of the Bogoliubov-de Gennes equations and the
Ginzburg-Landau analysis indicate that magnetization inevitably
appears when the two order parameters lead to a non-unitary
configuration and are spatially varying. We find that the
symmetry-breaking at the surface is unrelated to the chiral nature
of the bulk superconducting state, and therefore may be expected
in a much wider class of triplet superconductors. It would also be
very interesting to check whether similar effects occur at the
interfaces involving non-centrosymmetric superconductors, where
the singlet and the triplet components are intrinsically mixed in
the bulk yielding measurable spin effects at the
interface~\cite{Gorkov01,Vorontsov08,Sato09b, Schnyder12}, as well
as in the proximity structures with topological materials. We
leave this for future investigations.

{\it Acknowledgments.} This research has received funding from the
EU -FP7/2007-2013 under grant agreement N. 264098 - MAMA, and was
supported in part by US NSF via Grant No. DMR-1105339 (I.V.).

\end{document}

\title{Supplementary material for ``Magnetic intragap states and mixed parity pairing at
the edge of spin triplet superconductors''}

\author{Alfonso Romano}
\affiliation{CNR-SPIN, I-84084 Fisciano (Salerno), Italy and
Dipartimento di Fisica ``E. R. Caianiello'', Universit\`a di
Salerno, I-84084 Fisciano (Salerno), Italy}

\author{Paola Gentile}
\affiliation{CNR-SPIN, I-84084 Fisciano (Salerno), Italy and
Dipartimento di Fisica ``E. R. Caianiello'', Universit\`a di
Salerno, I-84084 Fisciano (Salerno), Italy}
\author{Canio Noce}
\affiliation{CNR-SPIN, I-84084 Fisciano (Salerno), Italy and
Dipartimento di Fisica ``E. R. Caianiello'', Universit\`a di
Salerno, I-84084 Fisciano (Salerno), Italy}
\author{Ilya Vekhter}
\affiliation{Department of Physics and Astronomy, Louisiana State
University, Baton Rouge, Louisiana, 70803, USA}
\author{Mario Cuoco}
\affiliation{CNR-SPIN, I-84084 Fisciano (Salerno), Italy and
Dipartimento di Fisica ``E. R. Caianiello'', Universit\`a di
Salerno, I-84084 Fisciano (Salerno), Italy}
%\date{\today}

\begin{abstract}
In this supplementary material to the main text we show that the
behavior of the magnetic field induced by the edge currents in a
chiral spin-triplet superconductor is very distinct from that of
the spin polarization, emphasizing the different origin of the
two. Furthermore, to elucidate this difference on the basis of the
Ginzburg-Landau approach, we numerically demonstrate that in a
triplet superconductor with a single component order parameter
that does not break time-reversal symmetry, the non-vanishing
magnetization at the edge of the system occurs because of the
non-unitary character of the mixed parity order parameter near the
boundary, while the edge currents are absent.
\end{abstract}

\maketitle

%%%%%%%%%%%%%%%%%%%%%%%%%%%%%%%%%%%%%%%%%%%%%%%%%%%%%%%%%%%%%%%

\section{Magnetic field induced by edge currents}

A chiral spin-triplet superconductor can sustain spontaneous
currents at the edge due to the time-reversal symmetry breaking in
the orbital channel. These currents, in turn, lead to the
appearance of a magnetic field near the boundary. One may ask
whether this field causes the spin polarization that we found in
the manuscript. The answer is negative, and below we show that the
spatial profile and the properties of the induced field are very
different from those of the spin-polarized states. Moreover, the
spin polarization persists even when there are no edge currents,
in agreement with our analysis.

We first compute the field due to the edge current. In principle,
this field is screened in the bulk on the scale of the London
penetration depth. However, in the immediate vicinity of the
interface the screening has little effect, and the field
distribution can be obtained directly from the spatial dependence
of the currents. For the two-dimensional slab configuration
analyzed in the paper, according to the Maxwell's law ${\bm
\nabla} \times {\bm B}=\mu_0 {\bm J_c}$, the magnetic field is
perpendicular to the plane ($z$-direction). At a distance $i_x$
from the edge it has an amplitude $B_{{\text{orb}}}^z(i_x)=-\mu_0
\sum_{j_x=1}^{i_x} J_c(j_x)$ with $\mu_0$ being the magnetic
permeability constant~\cite{Imai2012}.

\begin{figure}[!h]
\begin{center}
\includegraphics[width=0.42\textwidth]{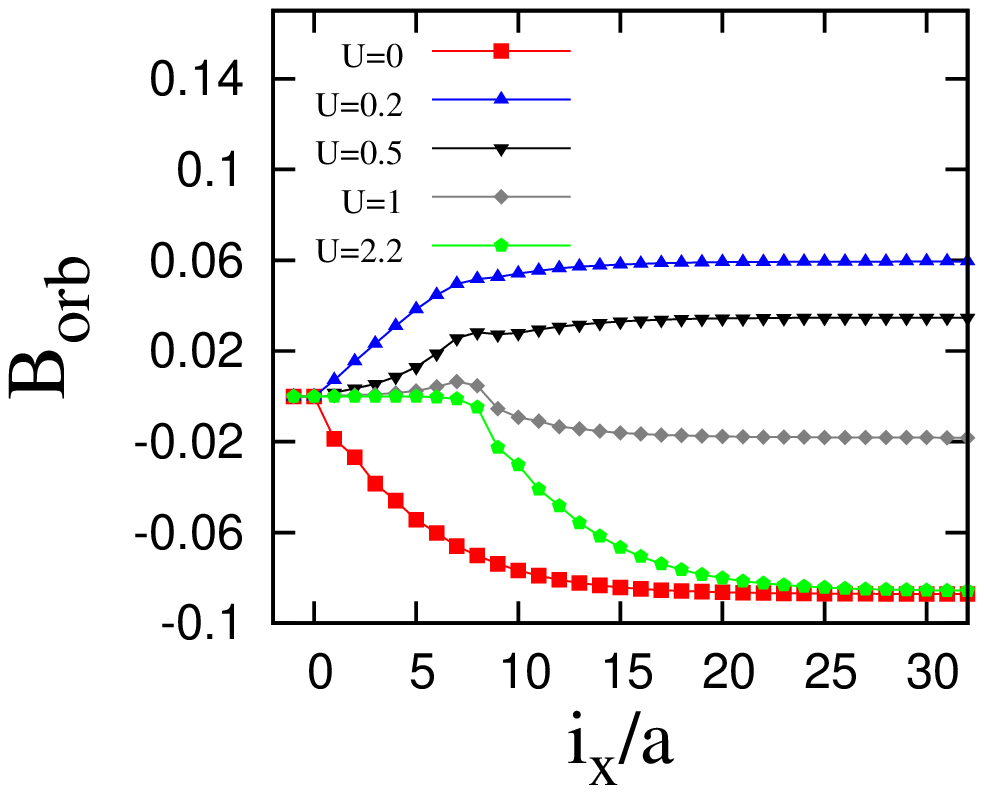}
\end{center}
\caption{(color online) Evolution of the magnetic field induced by
the edge currents for a chiral spin-triplet superconductor for
several values of the potential $U$. The scale of $B_{orb}$ is in
unit of $\frac{e\,t}{\hbar\,a}\mu_0$. We neglect here the Meissner
currents that affect the field distribution on the scale of the
London penetration depth, which is much greater than several
lattice spacings.} \label{fig:fig1}
\end{figure}

In Fig.\ref{fig:fig1} we show the behavior of $B_{{\text{orb}}}$
as a function of the potential at the edge. For $U=0$ there is an
induced magnetic field but no net spin magnetization, while above
the threshold both exist (see Fig. 2(b) in the main text). As
expected, due to the different mechanisms generating the spin
polarization and the orbital magnetic field, the spatial profile
of $B_{{\text{orb}}}$ is very different from that of the
magnetization. The spin-polarization is always of the same sign as
a function of the potential $U$ (see Fig. 2 in the main text), and
is peaked in the regime of the coexistence of the order parameters
of different parity. On the other hand, the orbital magnetic field
changes sign twice: around $U=0.2$, and then again  $U\geq 1$,
when the spin-polarization vanishes. The two are clearly of very
different origin.
%In the
%present analysis, we have neglected the Meissner screening that
%would have introduced counter currents suppressing this field on a
%length scale that is greater than the London penetration depth
%(i.e. in the inner side of the superconductor).

We note that the magnitudes of the field generated by both
mechanisms are comparable. To check this we convert the
spin-polarization into a magnetic field intensity~\cite{Imai2012}.
The spin-polarization yields a local magnetic field given by
$B_{{\text{pol}}}^z(i_x)=-\frac{\mu_0\mu_{B}}{a^2\,c}
(n_{\uparrow}(i_x)-n_{\downarrow}(i_x))$, where
$\mu_{B}=\frac{\hbar e}{2 m}$ is the Bohr magneton and $c$ the
lattice constant along the $z$ direction~\cite{Imai2012}. The
ratio between the prefactors of $B_{{\text{orb}}}$ and
$B_{{\text{pol}}}$ is
$r=(\mu_0\frac{e\,t}{\hbar\,a})/(\frac{\mu_0\mu_{B}}{a^2\,c}) \sim
t\, m (\frac{c}{a})$. Since the hopping amplitude is related to
the inverse of the effective mass $m^*$, for a generic filling the
ratio $r \sim O(1)$, thus implying that the two magnetic fields
are of the same order of magnitude. Hence, the field generated by
the spin polarization near the boundary will not have the same
profile as that due to the edge-currents.

\section{Spin-polarized edge states for single component
spin-triplet superconductors}

According to the symmetry constraints, only the $p_x$-component of
the spin-triplet order parameter appears in the Ginzburg-Landau
expansion (Eq. 2 of the main text). In principle, the term
$\partial_y \eta_y$ is also allowed~\cite{Kuboki:11}, but does not
contribute if the translational invariance along the interface
holds. Then, the time-reversal symmetry breaking by the bulk
chiral triplet state is not a necessary condition for the
occurrence of the magnetization of the Andreev bound states.
Indeed, the same result is obtained for purely real $p_x$ bulk
triplet superconductivity, while for the imaginary $p_x$ bulk
pairing with real subdominant $s$-wave pairing near the interface
no magnetization appears.

\begin{figure}[!h]
\begin{center}
\includegraphics[width=0.42\textwidth]{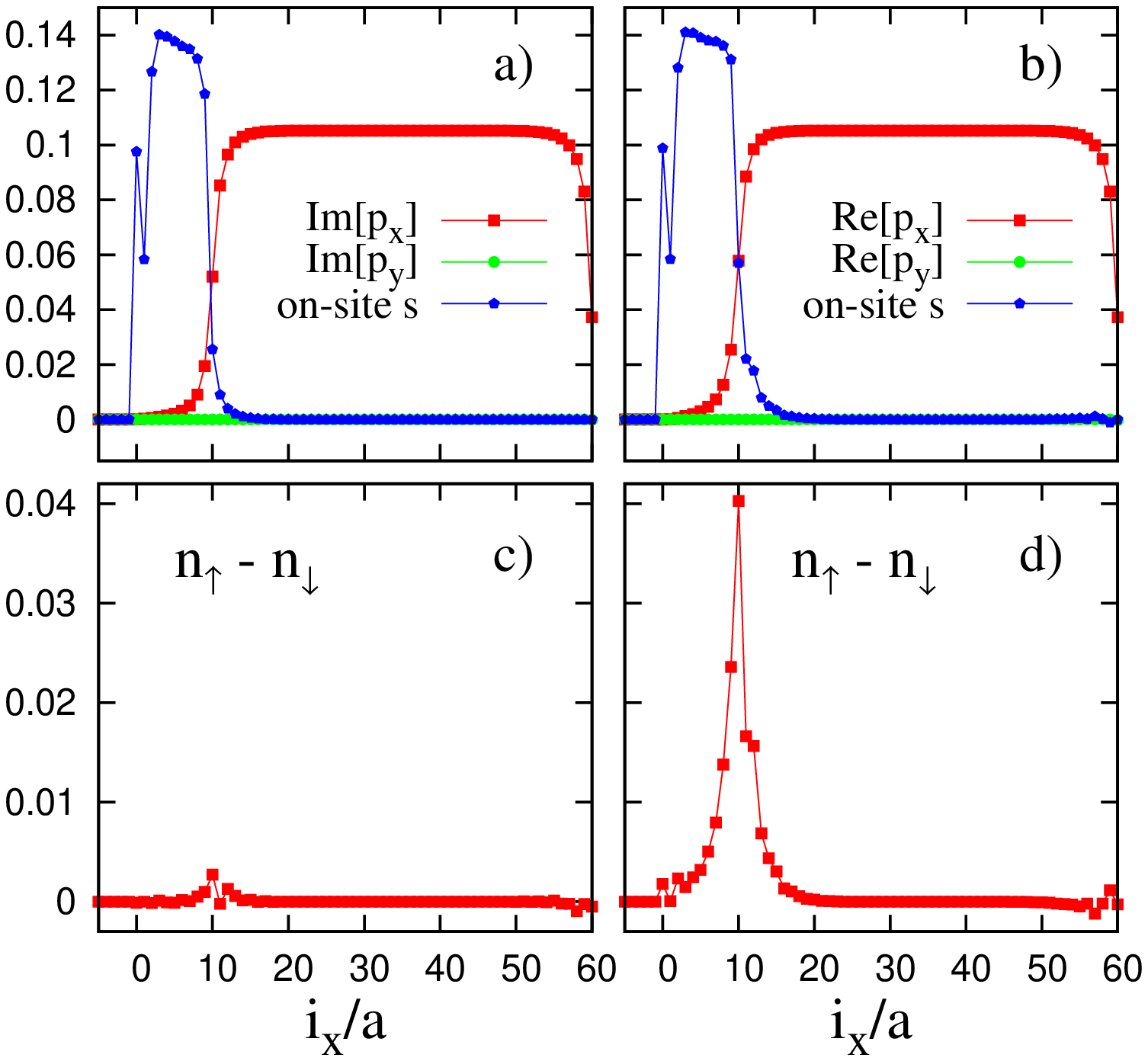}
\end{center}
\caption{(color online) Spatial evolution of singlet and triplet
order parameters and of the magnetization in the case of unitary,
a) and c), and non-unitary states, b) and d), respectively, as
realized in a thin vertical slab of the system around the point
($i_x=10$). A real spin singlet on-site pairing amplitude and a
purely real or a purely imaginary triplet pairing amplitude are
assumed in input ($i_x < 10$ and $i_x > 10$, respectively).}
\label{fig:fig2}
\end{figure}
%
%%%%%%%%%%%%%%%%%%%%%%%%%%%%%%%%%%%%%%%%%%%%%%%%%%%%%%%%%%%%%%%

To explore this connection, we consider a situation when in
proximity of the left edge ($0<i_x<\bar i_x$, with $\bar i_x=10$)
a purely local $s$-wave potential is the only source of pairing,
with $V$ still effective in the remaining part of the system, and
use as input a real $s$-wave order parameter for $0<i_x \le \bar
i_x$ and a purely imaginary $p_x$-wave one for $i_x>\bar i_x$.
Working at the same electron density as before (with $U=0$), and
without the self-consistent iterative procedure (which would
unavoidably lead to a mixing of real and imaginary components), we
see from the panels a) and c) of Fig.~\ref{fig:fig2} that, in
spite of the parity mixing occurring around $i_x=\bar i_x$, no
appreciable spin polarization is observed. On the other hand, when
in the same configuration one uses for $i_x>\bar i_x$ a purely
real $p_x$ input order parameter to generate a non-unitary mixed
phase around $i_x=\bar i_x$, a significant magnetization clearly
develops in the mixing region. This is exhibited in the panels b)
and d) of Fig.~\ref{fig:fig2}, and supports our conclusion based
on symmetry arguments.

In this case of a single component spin-triplet superconductor
there are no edge currents and thus the spin-polarization due to
the non-unitary character of the order parameter is the unique
source of magnetization at the boundary.